# Accelerated X-Ray Fluorescence Computed Tomography via Multi-Pencil-Beam Excitation

Ryder M. Schmidt[1], Daiki Hara[1], Jorge D. Vega[1, 2], Marwan B. Abuhaija[1, 2], Brett Bocian[1, 2], Wendi Ma[1], Nesrin Dogan[1], Alan Pollack[1], Ge Wang[3], John C. Ford[1], Junwei Shi[1]

1. Department of Radiation Oncology, University of Miami, Miller School of Medicine, Miami, FL, USA;
2. Department of Biomedical Engineering, University of Miami, Coral Cables, FL, USA
3. Department of Biomedical Engineering, School of Engineering, Rensselaer Polytechnic Institute, NY, USA

*Correspondence: (RMS) rms319@miami.edu; (JS) jxs1725@med.miami.edu

Abstract—X-ray fluorescence computed tomography (XFCT), a form of X-ray molecular imaging, offers detailed quantitative imaging capabilities for high-Z metal nanoparticles (MNPs), which are widely studied for their applications in multifunctional theranostics. Due to its affordability and accessibility, the benchtop XFCT prototype typically employs a single-pixel detector (SPD) with single-pencil-beam (SPB) X-ray excitation. While this design (resembling the first-generation CT geometry) achieves reliable detection sensitivity, it is hindered by long imaging times. The use of simultaneous multiple-pencil-beam (MPB) excitation presents a promising solution to significantly reduce imaging times. In this study, we developed a repeatable workflow that combines Monte Carlo (MC) simulations and 3D printing to design Nbeam-MPB collimator, where Nbeam is the number of beams generated by the collimator. As an initial test, we fabricated a 2-MPB collimator and evaluated the performance of 2-MPB-based XFCT imaging on a physical phantom and small animals surgically implanted with agarose pellets containing gold chloride (H[AuCl4]). The results demonstrated a 2x acceleration in image acquisition without compromising the contrast-to-noise ratio (CNR). We further investigated the concept of Nbeam-MPB acceleration on the MC computational XFCT system, which confirmed the feasibility of achieving at least 4x acceleration with 4-MPB excitation. Combined with additional system optimization, such as X-ray beam flux optimization, XFCT imaging could be further accelerated, reducing acquisition time from hours to minutes and meeting the requirements for routine MNP imaging.

# INTRODUCTION

X-ray fluorescence (XRF) analysis is a non-destructive analytical technique for determining a sample's elemental composition by using an X-ray source and an energy sensitive detector. The basic principle of operation relies on the detection of characteristic radiation produced through photoelectric interactions between incoming photons and the high-Z material within the sample. XRF analysis has been used for elemental analysis of biological samples for over 60 years [1]. X-ray fluorescence computed tomography (XFCT) uses the same operating principles as XRF analysis, with the addition of tomographic techniques to simultaneously depict the distribution and concentration of high-Z materials within a sample. Early XFCT imaging primarily utilized a monoenergetic pencil beam source generated by synchrotrons [2].

Recently, benchtop XFCT has become a reality. These systems typically use a collimated X-ray tube and an energy-sensitive single-pixel detector (SPD). The X-ray source is often collimated into a single pencil beam (SPB) and follows a first-generation computed tomography (CT) geometry, requiring multiple projections to reconstruct an image, with translation and/or rotation steps between each projection. At each projection, the collected X-ray spectra contain XRF peaks (i.e., characteristic radiation) from various elements in the beam path and/or sample, along with Compton-scattered background photons. This background energy spectrum can be fitted using various methods, such as a third-degree polynomial function, to extract the net XRF counts. By isolating the net XRF counts of the element of interest across different rotational and translational positions, a sinogram can be constructed for image reconstruction. Several reconstruction techniques are currently used, including algebraic methods such as filtered back projection (FBP) and iterative approaches like maximum likelihood expectation-maximization (MLEM) [3-5].

Several research groups have successfully developed benchtop XFCT systems for preclinical small animal research [6-12]. Many of these efforts focus on XFCT as a preclinical imaging tool for novel cancer therapies utilizing high-Z metal nanoparticles (MNPs) [13, 14]. In recent years, MNPs have garnered significant interest in cancer therapeutics, particularly in radiotherapy [15, 16]. These nanoparticles can locally enhance radiation doses, increasing DNA damage through higher photoelectric interactions and increased Auger electron production [17-19]. Among them, gold nanoparticles (AuNPs) are especially attractive due to their easily modifiable surface properties, relative chemical and biological stability, and straightforward fabrication [20, 21]. Additionally, other metal nanoparticles, such as gadolinium (GdNPs), Hafnium (HfNPs) platinum (PtNPs), have also been explored as novel radiosensitizers [22, 23].

XFCT is a promising technique for in vivo MNP imaging, providing both high spatial resolution and accurate quantification [1, 6, 14, 24]. It has proven valuable for enhancing the efficiency of small animal experiments through longitudinal in vivo imaging and is expected to play a crucial role in accelerating the clinical translation of novel MNP-based multifunctional cancer therapies [14, 25].

XFCT imaging requires an energy-dispersive detector, specifically a solid-state diode detector capable of distinguishing incident photon energies. Various SPDs and pixel array detectors (PADs) are available, utilizing different materials such as silicon drift detectors (SDDs), cadmium telluride (CdTe), cadmium zinc telluride (CZT), and germanium (Ge) detectors. For benchtop XFCT imaging, SDD- and CdTe-based SPDs are commonly used due to their affordability, high energy resolution, compact size, and ease of operation [12]. These detectors offer excellent energy resolution (~100–200 eV). SDDs are typically used for photon energies in the 0.1–30 keV range, while CdTe detectors are better suited for slightly higher energies, ranging from 10 to 200 keV. The choice of detector type (e.g., SDD vs. CdTe, SPD vs. PAD) directly influences the optimal excitation scheme.

XFCT imaging systems utilizing pencil beam, fan beam and cone beam excitation sources have been realized. Fan beam and cone beam XFCT imaging have the potential to improve imaging speed by stimulating a whole 2D or 3D plane, respectively. However, they require some collimation in front of the 1D- or 2D-PADs to identify the position information of XRF photons emitting from the object following excitation. Pencil beam excitation inherently provides position information for use in SPD-based XFCT imaging, with high resolution and sensitivity, but is also very time consuming, making it challenging for high-throughput in vivo studies.

It has been proven through in silico, in vitro and in vivo studies that PAD-based systems have the potential to reduce XFCT imaging time compared to SPD-based systems [7, 26-28]. For example, a MC study by Jones et al., published in 2011, demonstrated the feasibility of using polychromatic cone-beam photon sources for XFCT imaging [29]. Their findings showed that polychromatic cone-beam excitation allows for high quality image reconstruction with faster acquisition times compared to pencil beam excitation.

However, PAD-based XFCT systems typically suffer from compromised sensitivity due to reduction in useful signal caused by detector collimation. In a study by Moktan et al., which evaluated the differences between single pinhole collimation and parallel hole collimation for a 1D PAD, it was determined that there is a tradeoff between imaging sensitivity and image reconstruction quality, with the optimal detector collimation being specific to each case [30]. Jones et al. further explored the impact of multiple detectors on image quality through Monte Carlo simulations, concluding that increasing the number of detectors improved image reconstruction [29]. To enhance detection sensitivity in PAD-XFCT, Deng et al. developed a system incorporating multiple 2D PADs with multi-pinhole collimation, increasing XRF photon counts [31]. Despite these advancements, the high cost of PADs remains a significant barrier to the affordability and widespread adoption of PAD-XFCT systems. [26].

Unlike the PAD-XFCT systems, which rely on detector collimation to localize emitted XRF photons, the SPD-XFCT system determines positional information directly from the pencil beam excitation. Compared to the fan or cone beam excitation used in PAD-XFCT, the pencil beam generates a lower scattering background in the collected spectra. This combination of eliminating detector collimation and reducing scattering background results in higher detection sensitivity. However, pencil beam XFCT imaging requires significantly longer imaging times for object rotation and translation. Efforts can be made to accelerate image acquisition via rotation and/or translation modifications. Our group has explored the sparse-view strategy for SPD-XFCT acceleration by reducing the number of rotation angles

[32]. By using an L1-norm regularized expectation maximization (L1-EM) algorithm, we demonstrated a high-performance XFCT imaging with as few as 6 rotation views.

In this study, we explored another acceleration strategy for SPD-XFCT by implementing simultaneous multi-pencil-beam (MPB) excitation to reduce the number of translation steps. We developed a repeatable workflow to design and construct MPB source collimators using MC simulations and 3D printing. We successfully fabricated a 2-MPB collimator and performed phantom and small animal XFCT imaging experiments to validate this idea. We also investigate the potential of further acceleration by conducting MC simulations of Nbeam-MPB excitation XFCT imaging, where Nbeam indicates the number of pencil beams from a MPB collimator. To the best of our knowledge, this is the first time in which simultaneous multi-pencil beam excitation using a polychromatic benchtop x-ray tube for XFCT has been tested in any capacity.

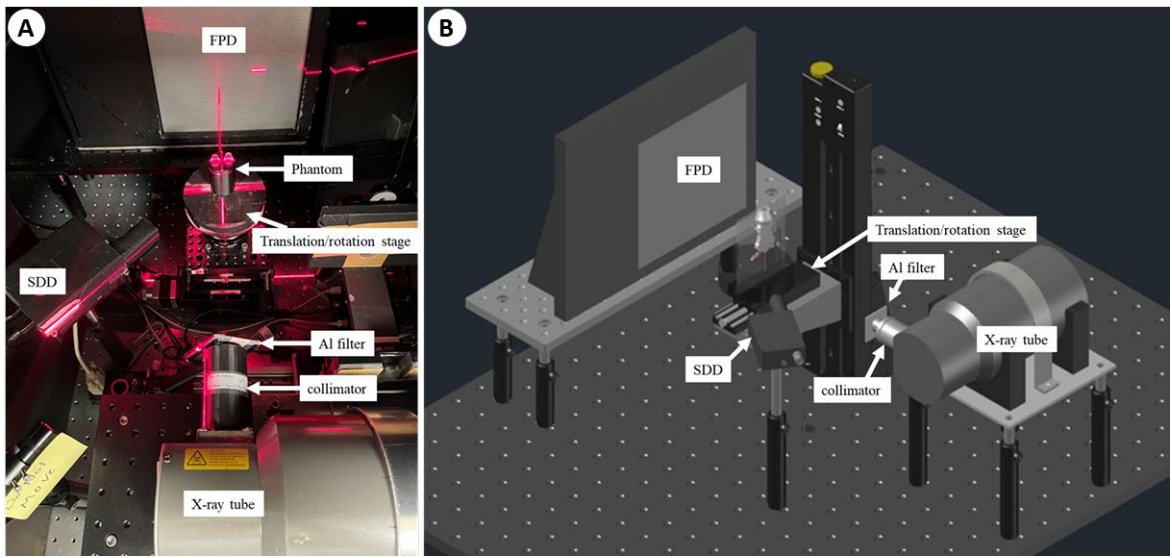

**Figure 1.** Small animal multi-modality imaging system (SAMMI) for multi-modal CT and XFCT. A) photography. B) AutoCAD 3D rendering.

# Methods

## Small Animal Multi-Modality Imaging (SAMMI): XFCT+CT

We have prototyped a SAMMI platform for multimodal CT and SPD-XFCT imaging [33]. The experimental setup and a 3D rendering of SAMMI are shown in Figures 1A and 1B, respectively. An X-ray tube (MXR 225/22, COMET AG, Flmatt, Switzerland) with a small focal spot (1.0 mm) and 0.8 mm thick inherent Be filter was used for both CT and XFCT imaging. The imaging object is placed on a X-Y-Z translation/rotation stage (Thorlabs, Newton, NJ, USA) with a source-to-rotation axis distance of 35.0 cm. A 3D printed iron SPB collimator was used for high-precision beam scanning for SPD-XFCT imaging. A silicon-based SPD or silicon drift detector (SDD, Amptek, Bedford, USA) with a 70 mm2 active area, 122 eV energy resolution at 5.9 keV, and maximum count rate of 4×106 sec-1, was placed 10 cm from the isocenter at 120° to the excitation beam. This SPD detector was operated at a bias voltage of -133 V with 4 µs peak time and 3x gain for collection of XRF photons. An amorphous silicon flat panel detector (FPD, PerkinElmer, Waltham, MA, USA) with a 20×20 cm2 active area and 200 µm pixel size (i.e., 1000×1000 pixels2) was placed behind the phantom perpendicular to the beam central axis to acquire transmission X-ray projection for CT, with a source-to-flat panel detector distance of 52.5 cm.

## Accelerated XFCT Imaging with MPB Excitation

As previously mentioned, single pencil beam XFCT follows 1st-generation CT geometry – meaning the pencil excitation beam has to make many translation steps, spanning the width of the phantom. For the SPB scanning technique, the number of linear translation steps ($N_L$) is determined by the diameter of the object being imaged ($D_{obj}$) and the linear step size ($L_s$):

$$N_L = (D_{obj}/L_s) + 1 , \qquad (1)$$

where NL is rounded up to the nearest whole number if needed. In the case of MPB scanning, the number of linear translation steps becomes:

$$N_L = \left(\frac{D_{obj}/L_s}{n}\right) + 1 , \qquad (2)$$

where n is the number of excitation beams and $N_L$ is rounded up to the nearest whole number if needed. The excitation geometry for SPB excitation, 2-MBP excitation, and 3-MPB excitation is shown in Figure 2A, 2B, and 2C, respectively.

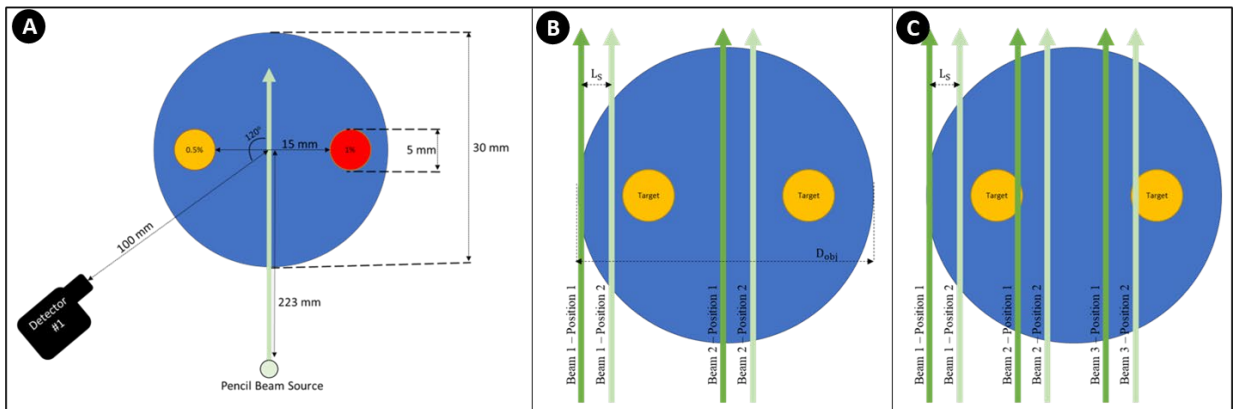

**Figure 2**: Diagram of SPD-XFCT imaging with SPB excitation (A) and N*beam*-MPB excitation (B: N*beam*=2; C: N*beam*=3). SPB: single pencil beam; MPB: multi pencil beam.

## MPB Collimator Design and Fabrication

To design the MPB collimator, MC simulations utilizing TOPAS were used to model the x-ray tube and test different collimator design (Figure 3). The x-ray tube was simulated to match the x-ray tube in our benchtop system mentioned earlier with a 1 mm diameter electron source at 64 kVp, a tungsten anode at 20°, a 2.45 mm diameter window with 0.8 mm inherent Be filter. The collimator was simulated using Cerrobend material. Cerrobend is a brand name for a material known as Wood's metal or Lipowitz's alloy, (50% Bi, 26.7% Pb, 13.3% Sn and 10% Cd, ρ=9.4 g/cm3), a highly attenuating material with a relatively low melting point (70℃) – used clinically for electron beam shaping. The base of the collimator was 4.5×4.5×1.3 cm3, with a 3.3 cm diameter and 1.3 cm height cylindrical cutout. The collimator nozzle was 7.6 cm in length (total collimator length = 8.9 cm), and 4 cm in diameter with two pinhole cutouts 1 mm in diameter with angles of ±1.27° from the collimator central axis. The simulations were broken up into two steps for a faster workflow, utilizing a phase space placed on the distal side of the Be filter. Through this MC beam collimation simulation, we can assess the designed collimator's dimensions and material properties, making adjustments as needed.

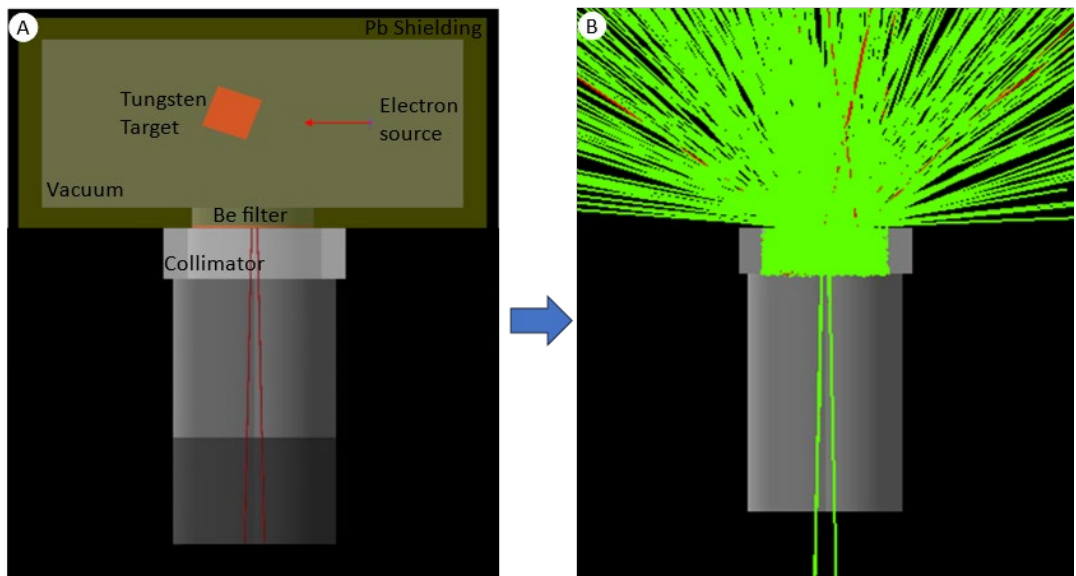

**Figure 3**: Screenshot of TOPAS MC X-ray tube used to determine feasibility and aid in the design of MPB collimator. Simulated components include the electron source directed at a tungsten anode in vacuum, with lead shielding and a Be window, and a MPB collimator made of Cerrobend. Note the 2-MPB paths are shown in red on the left for better visualization. Simulations were broken up into two steps, with a phase space placed at the distal end of the Be filter. Green beams indicate X-ray photons.

Following the MC simulations, a mold used for Cerrobend casting was designed using AutoCAD (AutoDesk Inc., San Rafael, CA, 2023), and 3D printed with an Ultimaker3 3d printer (Ultimaker BV, Utrecht, Netherlands, 2021) using PC Black filament, which has a suitably high melting point (>110°C). The AutoCAD design and 3D printed mold did not include the 1 mm diameter pinhole paths, instead, holes were printed in the inner and outer mold and brass tube inserts were used to block Cerrobend from filling these beam paths. After the two 3D printed mold pieces were assembled and the brass tube was put in place, the mold was filled with liquid Cerrobend through the large hole atop the outer mold. The inner mold piece, brass tubes, and some of the outer mold around the base were removed after the Cerrobend hardened. After constructing the dual-pencil beam collimator, it was tested by measuring the beam intensity and full-width at half-maximum (FWHM) of each beam on the FPD. Using the imaging geometry and trigonometry, the FWHM at the imaging isocenter was calculated. To test the concept of MPB-based XFCT imaging, a dual pencil beam (2-MPB) collimator was designed and constructed in this study.

## 2-MPB Phantom Experiments

After the creation and testing of the customized 2-MPB collimator, experimental XFCT imaging of a gold-containing phantom was done. The experimental setup of the benchtop XFCT imaging system is shown in Figure 1. The X-ray tube was operated with a voltage of 64 kVp and current of 10 mA to maximize gold (Au) fluorescence production and match the parameters of the MC simulations. An additional 2-mm thick Al filter was used following collimation to suppress the L-edge fluorescence from the tungsten target of the X-ray tube and attenuate low energy photons (<Au L-shell binding energy) that would only serve to add additional dose and background signal.

Commercially available gold chloride (MilliporeSigma) was used to create two gold-containing targets within the phantom. The phantom was 3 cm in diameter and 4.5 cm in height, with the two Au-targets at 7 mm diameter and 1.0 and 0.5 wt.%, receptively. The imaging setup was as described in part A of the methods in this paper.

The number of translation steps per rotation was reduced from the SPB scanning as we did previously [3], from 21 to 11, keeping the 1.5 mm/step size; and the number of rotation steps was kept at 36 with 10°/step size, totaling 396 projections, following Equation (2). Each projection used a beam on time of 30 seconds. At each projection Au-$L_{\beta 1}$ net XRF counts over the energy range of 11.44±0.35 keV were extracted by curve fitting the Compton background using an adaptive iteratively reweighted penalized least-squares (airPLS) curve fitting algorithm. Details of this iterative polynomial curve fitting method can be found in [34-36].

## 2-MPB Small Animal Experiments

Following the XFCT studies using a gold-target containing phantom, in vivo post-mortem mouse models were used to further characterize/validate the imaging abilities of the 2-MPB XFCT imaging system. The same imaging set-up used for the phantom studies was used for these post-mortem mouse models including x-ray tube, collimator, filter material and thickness, rotation/translation stage, flat panel detector, silicon drift detector etc. Additionally, the same x-ray tube energy and current, beam on time, rotation and translation steps, and silicon drift detector settings were employed. The object being imaged, however, is now a post-mortem nude mouse, approximately 3 cm in diameter and approximately 10 cm in length. The mouse was first euthanized carbon dioxide gas and subsequently a gold target was implanted in the abdominal region of the mouse. The gold target was a cylinder approximately 10 mm in length and 5 mm in diameter, with a concentration of gold at 1.0 wt.%. A second mouse was prepared in the same manner using a gold target at 0.5 wt.%. The mouse was then placed in a 3 cm diameter Eppendorf tube and the remaining air was filled with formaldehyde. The Au XRF signal extraction follows the same procedure as described in the phantom studies.

## Nbeam-MPB Based XFCT Reconstruction

A single pixel within the measured sinogram ($P_i$) can be mathematically represented using

$$P_i = \sum_j W_{i,j} \cdot X_j, \tag{3}$$

where $W_{i,j}$ is the system matrix (also known as a forward matrix), $X_j$ is the pixelized object being imaged, $i$ is a single projection, and $j$ is a pixel within the object. Several reconstruction techniques have been developed to solve this reconstruction problem including algebraic, iterative and machine learning based [3-5]. To employ these reconstruction techniques, we first need to develop the system matrix ($W_{i,j}$), calculated by:

$$W_{i,j} = d_{i,j} \cdot e^{-\mu_{ex} \cdot l_{ex}} \cdot e^{-\mu_f l_f}, \tag{4}$$

where $d_{i,j}$ is the probability that an XRF photon is emitted from pixel $j$ excited by the beam $i$, and was calculated as the area of pixel $j$ covered by the excitation beam $i$, with total coverage equal to 1. The second and third term in Equation (4) accounts for attenuation of the excitation beam from source-to-pixel, and attenuation from pixel-to-detector, respectively. $\mu_{ex}$ is the linear attenuation coefficient of water at the excitation beam mean energy of 21 keV (0.075/mm), and $l_{ex}$ is the path length traveled by the excitation beam in water from the source to pixel $j$; $\mu_f$ is the linear attenuation coefficient of water at $L_{\beta 1}$ emission energy of ~12 keV (0.347/mm), and $l_f$ is the path length traveled in water from pixel $j$ to the detector.

The goal of XFCT reconstruction is to obtain Au distribution X from measurement P based on Equation (3), however, no direct solution exists due to the ill-condition of the system matrix W. To solve this reconstruction problem, a joint L1 and total variation (TV) regularized algorithm (L1-TV) was used. L1 regularization was used to reduce image artifacts and TV regularization was used to preserve the target shape – more details on this reconstruction algorithm can be found in the previous publication from our group [3].

For MPB reconstruction problems, the same workflow was followed as done for SPB, and the system matrix for 1-, 2-, 3-, and 4-MPB excitation reconstructions were all built following equation 4. However, as the number of beams increased (from 1 to 4), fewer projections were used and d_(i,j) values were adjusted to reflect pixels that were in the beam path and thus had a non-zero probability of generating XRF photons.

For all beam configurations dose to the entire phantom and to each Au-target was recorded. All reconstructed images were evaluated using contrast-to-noise ratio (CNR) for each ROI. CNR can be calculated using:

$$CNR = \frac{\mu_{ROI} - \mu_{BG}}{\sigma_{BG}}, \quad (5)$$

where $\mu_{ROI}$ is the average pixel intensity of a gold target, $\mu_{BG}$ is the average pixel intensity of the background, and $\sigma_{BG}$ is the standard deviation of the background. Because the phantom used has two targets CNR is calculated twice, once for each target. Following the 'Rose Criterion,' CNR>4 was considered easily distinguishable from background. All data processing and analysis was done using MATLAB (MathWorks, Inc., Natick, Massachusetts, version R2022b).

## Nbeam-MBP Monte Carlo Simulations

To further investigate the feasibility of the concept Nbeam-MPB-XFCT with Nbeam>2, we conducted a series of MC simulations of XFCT under 1-, 2-, 3-, and 4-MPB excitations, using the TOPAS toolkit [37]. Simulations were done using the geometry of the in-house custom-built SAMMI (Figure 1) used for phantom and in vivo experiments. These simulations utilized a 30 mm diameter phantom and two 5 mm diameter Au-targets at 0.5 and 1 wt.%, respectively. SpekCalc was used to generate the X-ray spectrum of 64 kVp, passing through a 0.8-mm Be inherent filter and a 2-mm Al additional filter (Figure 4D). This X-ray tube was simulated as a 2 mm diameter pencil beam source(s) with 0.1° angular spread. SpekCalc code has been validated with experimental and MC measurements and has been used by many groups to quickly and accurately acquire x-ray tube energy spectrums [38, 39]. A single-pixel CdTe detector was simulated with 25 mm2 active area composed of 47% Cd and 53% Te with density of 5.85 g/cm3, and 0.1 mm Be filter – placed 120° from the incident beam path at 10 cm from the imaging isocenter. All components were placed in a 6×7×11 cm3 air world.

Each imaging procedure used 1.5 mm linear translation step-sizes and 10° rotational step-sizes. In the case of multiple source beams, sources were placed with spacing of Dobj/n, where Dobj is the diameter of the object being imaged and n is the number of pencil-beam sources, with the first beam placed so its path hits the edge of the phantom (Figure 2), following the logic of equations 1 and 2 laid out in part B of the methods. Simulations were done with 500 million histories generated by each beam. As the number of beams increased, the number of histories simulated for each beam stayed constant, however, the total number of histories per projection (summed from each beam) increased. Screenshots of the MC simulated geometry for 1, 2, 3, and 4 pencil beam excitation sources with 1000 histories sampled from each beam are shown in Figure 4B, 4C, 4E, 4F, respectively. The collected energy spectrums of each detector for each projection were curve fitted using airPLS curve fitting algorithm to extract Au $L_{\beta1}$ net-XRF counts over the energy range of 11.44±0.35 keV. Imaging dose to each Au-target and to the entire phantom were scored during these MC simulations utilizing a built-in dose scoring method in TOPAS called 'DoseToMedium."

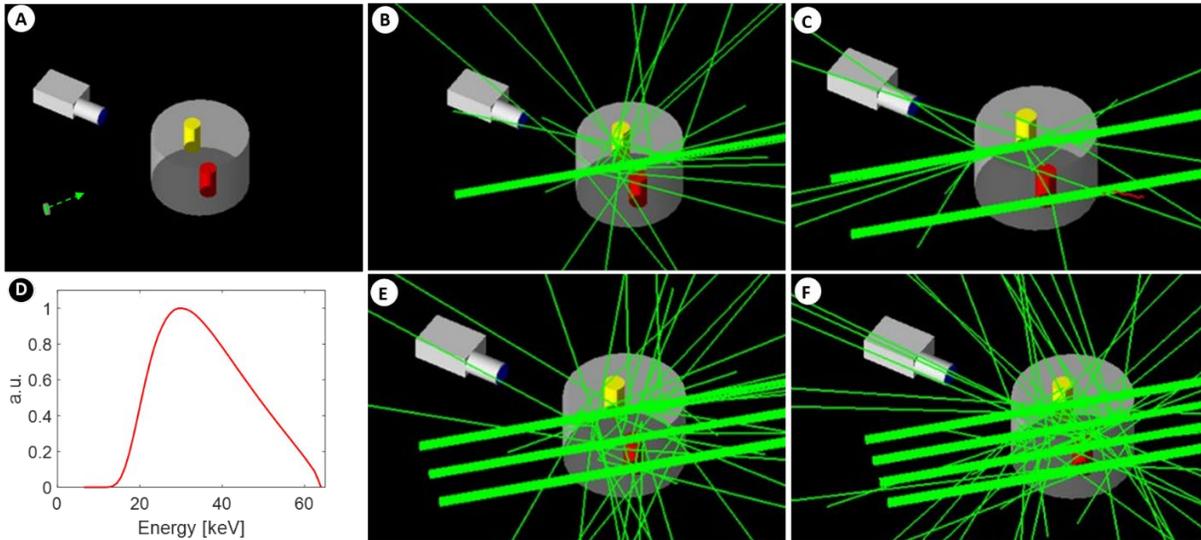

Figure 4: A) Screenshot of Monte Carlo Simulated geometry showing the x-ray source (green) directed at the phantom containing two gold-targets. The detector is shown in grey on the left of the phantom. B) Screenshot of Monte Carlo simulated single pencil beam XFCT with 100,000 simulated particles. C) Screenshot of Monte Carlo simulated two pencil beam XFCT with 100,000 simulated particles. D) Plot of energy spectrum generated via SpekCalc used as the excitation pencil beams. E) Screenshot of Monte Carlo simulated three pencil beam XFCT with 100,000 simulated particles. F) Screenshot of Monte Carlo simulated four pencil beam XFCT with 100,000 simulated particles.

# Results

## 2-MPB collimator fabrication

We fabricated a 2-MPB collimator by 3D-printing the mold and casting it with liquid Cerrobend. Figures 5A and 5B show the AutoCAD models of the outer and inner components of the collimator mold. Two small holes with diameters of 3 mm were designed to anchor brass tubes with an outer diameter of 3 mm and inner diameter of 2 mm. A larger hole was used for Cerrobend casting. The final fabricated collimator is shown in Figure 5C. To verify the precise alignment of the pinholes with the center of the X-ray tube focal spot, the pencil beams passing through the collimator were projected onto the flat-panel detector (FPD) in the SAMMI system (Fig. 1). The resulting X-ray projection image (Fig. 5D) exhibits sharp beam edges and high target-to-background contrast, confirming the successful fabrication and alignment of the collimator. Beam profile analysis determined that the full-width at half-maximum (FWHM) of each beam at the isocenter (center of the rotation stage) was 2.2 mm, while the distance between the two beam centers at the isocenter was 15.8 mm. This configuration is well-suited for XFCT imaging of mice with a diameter of ~3 cm.

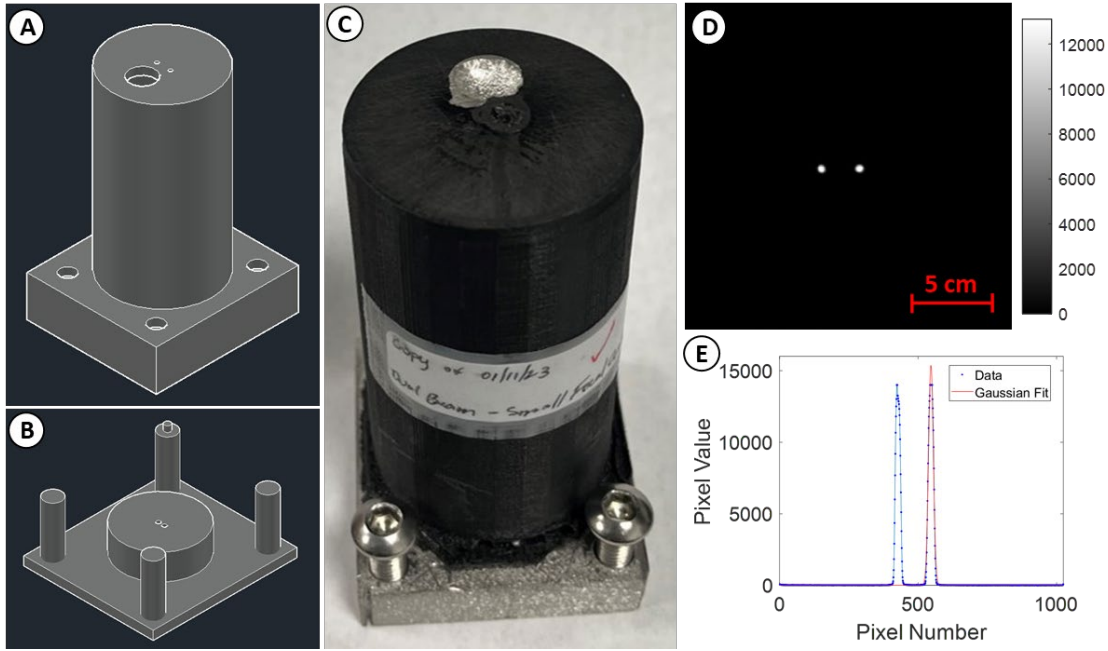

**Figure 5**: AutoCAD mold design for MPB collimator casting, with A) outer piece and B) inner piece. C) Fabricated 2-MPB collimator. D) Two pencil beams projected on the flat panel detector (FPD).

## 2-MPB XFCT Imaging: Phantom

We verified 2-MPB XFCT imaging on a small-animal-size phantom (diameter: 3 cm), which contains two gold [Au] targets (gold chloride) at 1.0 wt.% and 0.5 wt.% (Figure 6A). CT transverse slice was shown in Figures 6B, as a benchmark against 2-MPB XFCT imaging. Figure 6 C shows the sinogram of 2-MPB XGCT scanning with 11 translation steps and 36 rotation steps. Figure 6D shows the 2-MPB reconstruction overlapped with CT. The agreement between XFCT recovered targets with CT recovered targets demonstrates the localization accuracy of 2-MPB scanning strategy. The quantification accuracy of 2-MPB XFCT can be reflected on the accurate intensity ratio (2:1) for the Au targets with concentrations of 1.0 wt.% and 0.5wt.%.

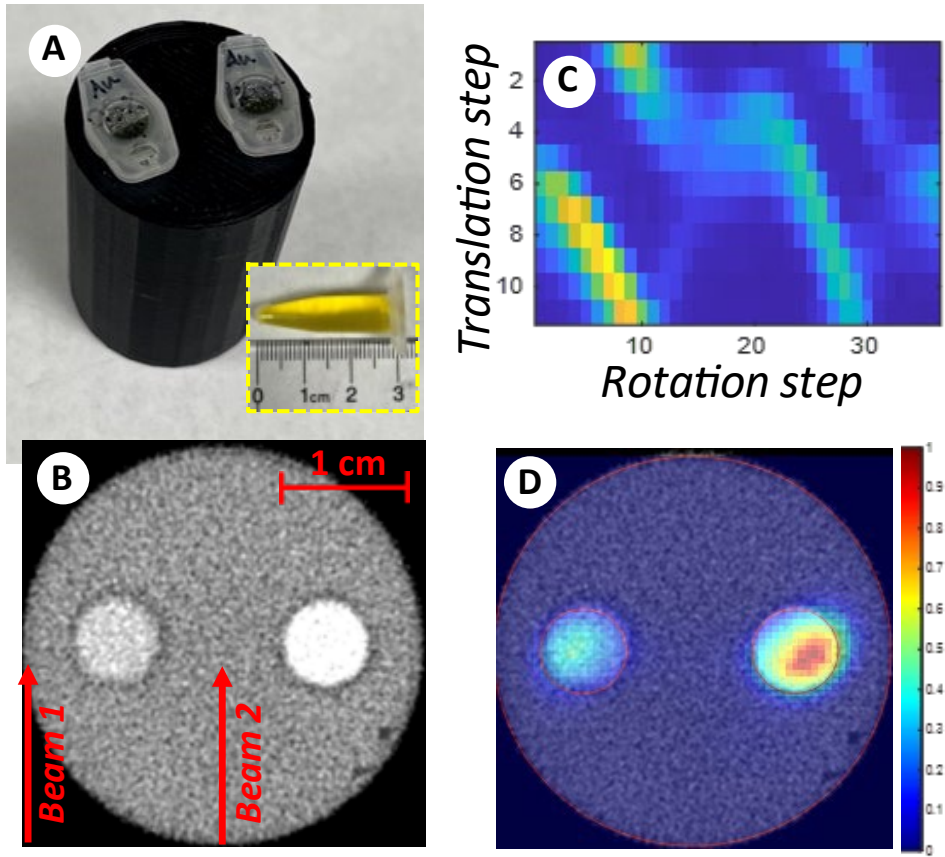

**Figure 6**: 2-MPB XFCT imaging. A) Small phantom containing Au-targets and a sub-image depicting the Au-targets. B) Axial CT slice through the phantom showing the two Au-targets with the two beamlines shown in red. C) Sinogram collected following 2-MPB excitation of this phantom. D) Overlay of the reconstructed XFCT and CT.

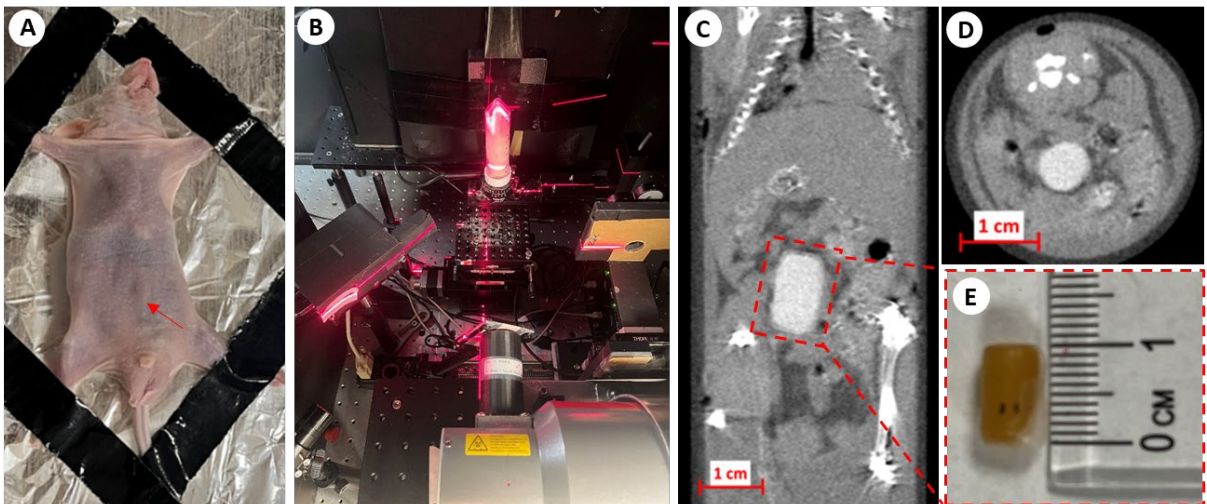

**Figure 7**: A) Post-mortem nude mouse, with red arrow depicting where the Au pellet was implanted. B) SAMMI system with the mouse placed on the rotation stage. C) and D) Coronal and transverse CT slices, highlighting the implanted E) Close up view of implanted Au-pellet prior to implantation.

## 2-MPB XFCT Imaging: Small Animal

A mouse model with an implanted Au-target is shown in Figure 7. With the mouse pre-implant shown in Figure 7A, the SAMMI system with the mouse on the rotation stage shown in Figure 7B, a coronal and axial CT slice of the mouse contain the Au-target is shown in Figures 7C and 7C, respectively, and a close-up view of the Au-target shown in Figure 7E.

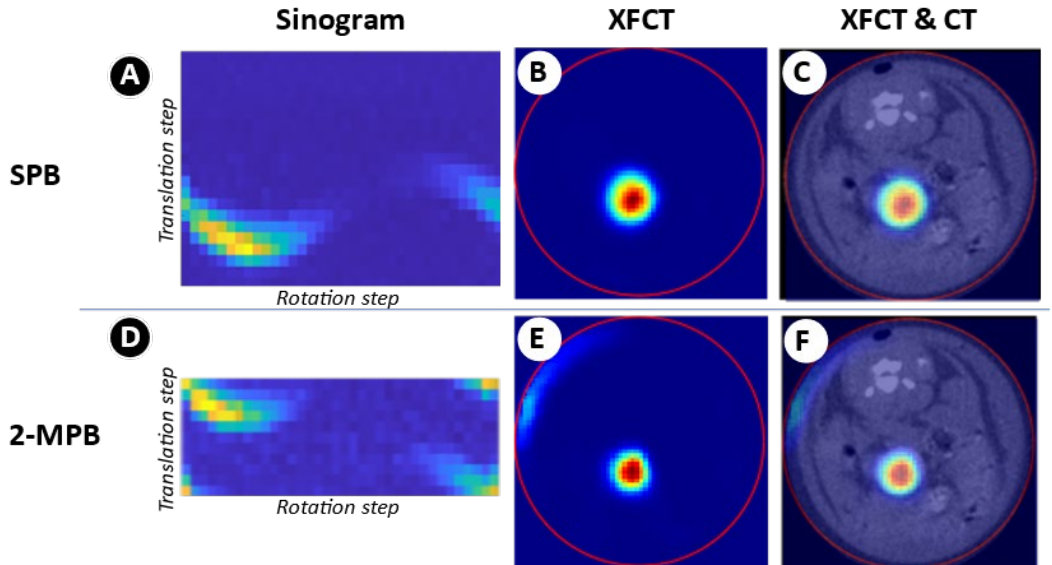

**Figure 8**: SPB-XFCT (top row) vs 2-MPB-XFCT (bottom row) imaging of small animal surgically implanted with 1 wt.% Au-pellet. A) Sinogram of SPB excitation. B) Reconstructed XFCT image utilizing SPB excitation. C) Overlay of axial CT slice with reconstructed XFCT image for SPB. D) Sinogram of 2-MPB excitation. E) Reconstructed XFCT image utilizing 2-MPB excitation. E) Overlay of axial CT slice with reconstructed XFCT image for 2-MPB.

Figure 8 compares SPB and 2-MPB XFCT imaging in a mouse surgically implanted with a 1.0 wt.% Au-pellet. Due to the twofold reduction in translation steps, the sinogram size in the 2-MPB case (Figure 8D) is compressed to half that of the SPB case (Figure 8A). The successful reconstruction of 2-MPB XFCT further demonstrates the advantages of the MPB strategy in accelerating imaging without compromising quality. A similar small animal experiment was conducted using a pellet with a lower Au concentration (0.5 wt.%), yielding comparable results, though not shown here. The color bar for each XFCT image was normalized to the maximum pixel intensity.

## Nbeam-MPB XFCT Imaging: Monte Carlo Simulations

Nbeam-MBP strategy with Nbeam>2 was investigated in our MC computational XFCT model. Collected sinograms from Nbeam-MBP excitation for 1, 2, 3, and 4 pencil beam sources is shown in Figures 9A-9D, respectively. Each sinogram was normalized to the maximum counts for the given excitation scheme for easier viewing. The MPB strategy can be conceptualized as a "sinogram fold," where projections from Nbeam simultaneous excitations are merged, effectively compressing the sinogram by a factor of 1/Nbeam in the translation dimension.

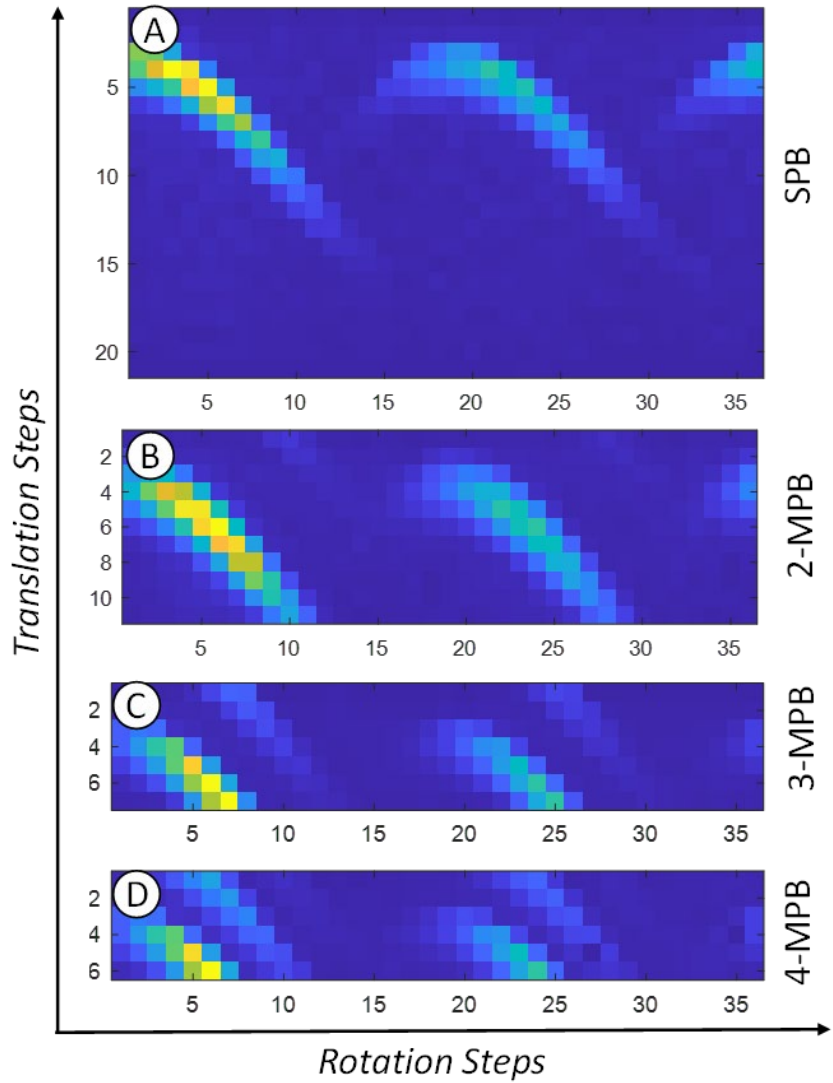

**Figure 9**: Sinograms constructed from SPD-XFCT under SPB, 2-, 3-, and 4-MPB excitation, respectively. All sinograms were normalized to its maximum pixel value.

Using sinograms generated from MC MPB excitation (Figure 9), the system matrix developed using the previously described method, and the joint L1-TV regularization reconstruction algorithm, XFCT images were reconstructed for the SPB, 2-, 3-, and 4-MPB schemes, as shown in Figure 10A, 10B, 10C, and 10D, respectively. Both ROI-1 (1.0 wt.%) and ROI-2 (0.5 wt.%) were clearly distinguishable with high target-to-background contrast, as confirmed by the Rose Criterion (Table 1). The recorded dose to the phantom and to each target (ROI-1 and ROI-2), are also presented in Table 1. As expected, dose per projection increased in a linear fashion (R2=0.976), while dose to the phantom as a whole and to each individual target had non-statistically significant changes ($p<0.05$).

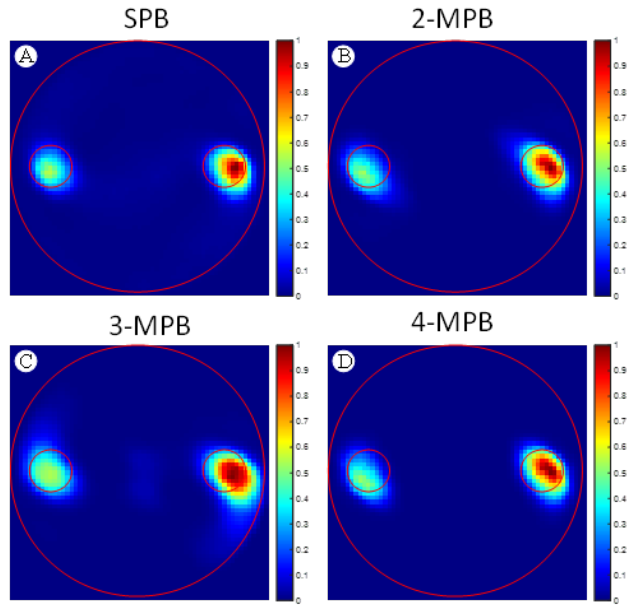

**Figure 10**: XFCT reconstructions in the scenarios of SPB (A), $N_{beam}$-MPB (B: $N_{beam}=2$; C: $N_{beam}=3$; D: $N_{beam}=4$) excitation, using the sinograms as shown in Figure 9.

**Table 1**: CNR and relative dose to each Au-target (ROI1: 0.5 w.t.[Au]% and ROI2: 1 w.t.[Au]%), and the whole phantom, for MC computational XFCT under SPB, $N_{beam}$-MPB ($N_{beam}=2$, 3, and 4) excitation schemes.

| Pencil beam excitation scheme | CNR | | Relative Dose | | |
|---|---|---|---|---|---|
| | ROI 1 (1.0 wt.%) | ROI 2 (0.5 wt.%) | Whole Phantom | ROI 1 (1.0 wt.%) | ROI 2 (0.5 wt.%) |
| 1 beam | 22.1 | 18.4 | 1.00 | 1.00 | 1.00 |
| 2 beams | 16.0 | 13.8 | 1.02 | 1.01 | 0.99 |
| 3 beams | 15.8 | 12.7 | 1.02 | 1.02 | 1.02 |
| 4 beams | 13.3 | 11.0 | 1.09 | 1.01 | 1.04 |

## Discussion

In this study, we proposed a novel acceleration strategy for SPD-based XFCT, aimed at developing a cost-effective system to support widespread adoption for routine in vivo imaging in the MNP theranostics community. Specifically, we implemented a multi-pencil beam (MPB) approach to simultaneously excite object at multiple positions, considerably reducing the translation scanning steps compared to the traditional single-pencil beam (SPB) based SPD-XFCT. We validated the MPB concept though MC simulations using TOPAS, successfully customized a 2-MPB collimator, and evaluated its performance through phantom and small animal experiments. The experimental results demonstrated that SPD-XFCT with the 2-MPB collimator achieved a 2× acceleration in imaging without compromising localization accuracy or target-to-background contrast. In silico MC simulation alludes the 4× or even higher acceleration. Theoretically, MPB strategy will not increase the imaging dose, provided no overlap occurs during the translation scanning. This approach represents a significant step toward advancing efficient and scalable imaging solutions for MNP-based theranostics.

It is worth noting that the 2-MPB collimator developed in this study was specifically designed for the 1 mm focal spot of our COMET MXR-225-22 X-ray tube, which features a dual focal spot mode with different output powers (1 mm, 640 W; 5.5 mm, 3000 W). Due to the high X-ray attenuation properties of Cerrobend, our collimator design can be readily adapted for the 5.5 mm focal spot mode with minor adjustments, such as aligning the MPB pinholes to the lager focal spot center, enabling acceleration of over 3×. Therefore, combining a 4-MPB configuration with a higher-flux X-ray beam has the potential to achieve more than 12× acceleration, reducing the single-slice XFCT imaging time from the current ~3 hours to approximately 15 minutes. This advancement would make the technique highly suitable for routine applications.

As illustrated in the mechanism (Fig. 2), the sinogram size is compressed in MPB-based scanning, which may affect the ill-posedness of the XFCT reconstruction problem. While the traditional MLEM algorithm used in this study demonstrated satisfactory performance, advanced regularized reconstruction methods and AI-based learning algorithms have the potential to further enhance imaging quality in MPB-based accelerated XFCT. Additionally, employing multiple SPDs positioned at different angles relative to the X-ray beam's central axis can provide complementary information, improving XFCT reconstruction accuracy. The inclusion of such complementary data also has the potential to enhance the detection sensitivity of XFCT imaging. We plan to utilize our established MC computational XFCT model to investigate the optimal arrangements of multiple SPDs.

In this study, we used gold chlorides to evaluate the feasibility of the MPB technique for noninvasive monitoring of metal agents or MNPs. As a novel spectroscopic molecular imaging approach, the developed method can be adapted to image MNPs composed of various high-Z metal materials, including gadolinium ($Z = 64$), hafnium ($Z = 72$), bismuth ($Z = 83$), lead ($Z = 82$), and others. Future efforts will focus on optimizing the excitation spectra for diverse or multipixelated MNP imaging using the proposed MPB techniques.

# Conclusion

This study introduces a novel multiple-pencil-beam excitation strategy for XFCT imaging that successfully addresses the longstanding challenge of long acquisition times. Through a systematic evaluation consisting of Monte Carlo simulations, small phantom studies, and in vivo imaging, we demonstrated the feasibility and advantages of this approach. Our findings pave the way for future developments of cost-effective XFCT technology, with the potential to accelerate its adoption in routine preclinical research, spurring further clinical translation of novel cancer treatments based on multi-functional metal nanoparticles.